\documentclass[usegraphicx,usenatbib]{mn2e}

\usepackage[a4paper,total={17.8cm,24.0cm},centering]{geometry}
\usepackage{times}

\newcommand{\dd}{\mathrm{d}}
\newcommand{\mnras}{MNRAS}       
\newcommand{\aap}{A\&A}

\newcommand{\refeq}[1]{equation~(\ref{#1})}

\title[Anisotropic axisymmetric Jeans models]
{Anisotropic Jeans models of stellar kinematics: second moments including proper motions and radial velocities}

\author[M.~Cappellari]{Michele Cappellari\thanks{E-mail:
cappellari@astro.ox.ac.uk}\\
Sub-Department of Astrophysics, Department of Physics, University of Oxford, Denys Wilkinson Building, Keble Road, Oxford, OX1 3RH}

\date{29 November 2012}

\pagerange{\pageref{firstpage}--\pageref{lastpage}} \pubyear{2012}

\begin{document}
\label{firstpage}
\maketitle

\begin{abstract}
This is an addendum to the paper by Cappellari (2008, MNRAS, 390, 71), which presented a simple and efficient method to model the stellar kinematics of axisymmetric stellar systems. The technique reproduces well the integral-field kinematics of real galaxies. It allows for orbital anisotropy (three-integral distribution function), multiple kinematic components, supermassive black holes and dark matter. The paper described the derivation of the projected second moments and we provided a reference software implementation. However only the line-of-sight component was given in the paper. For completeness we provide here all the six projected second moments, including radial velocities and proper motions. We present a test against realistic N-body galaxy simulations.
\end{abstract}


\begin{figure*}
  \includegraphics[width=0.85\textwidth]{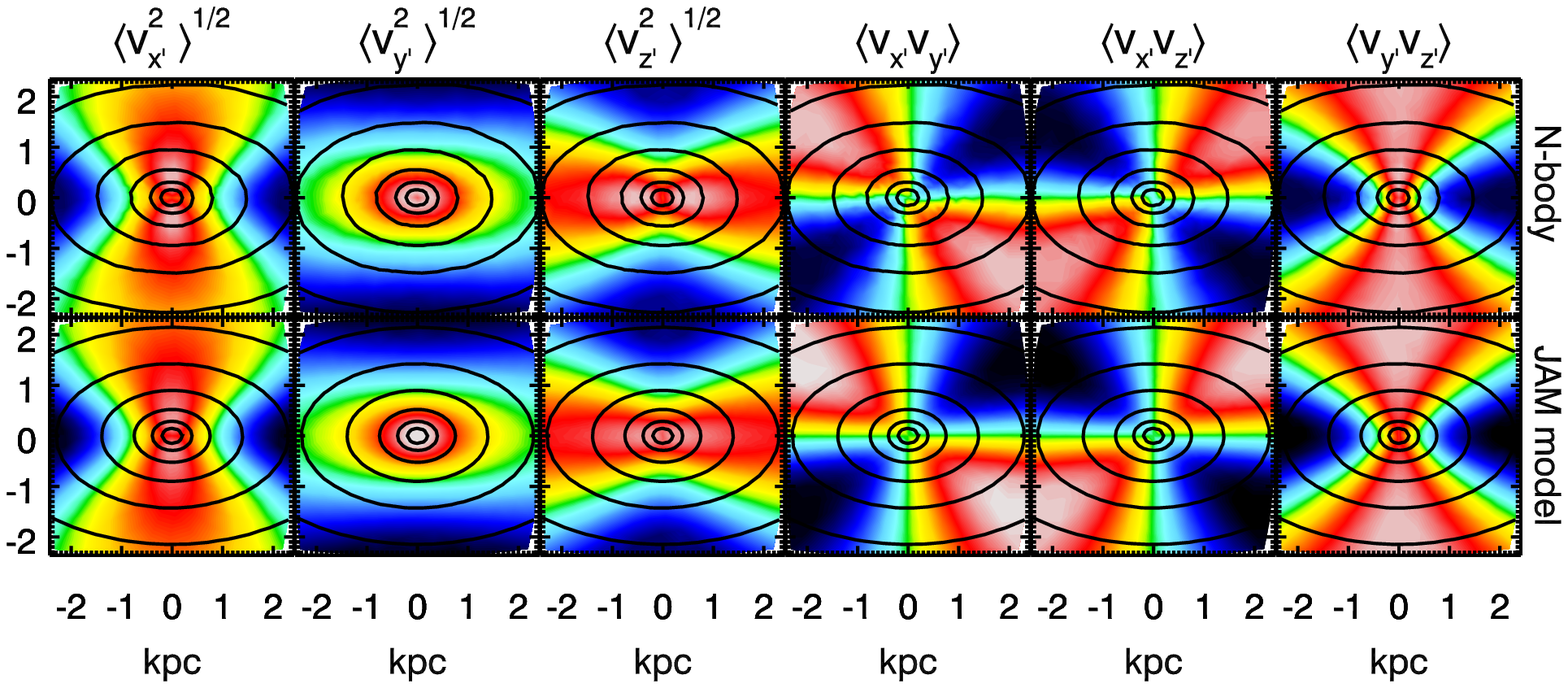}
  \includegraphics[width=0.85\textwidth]{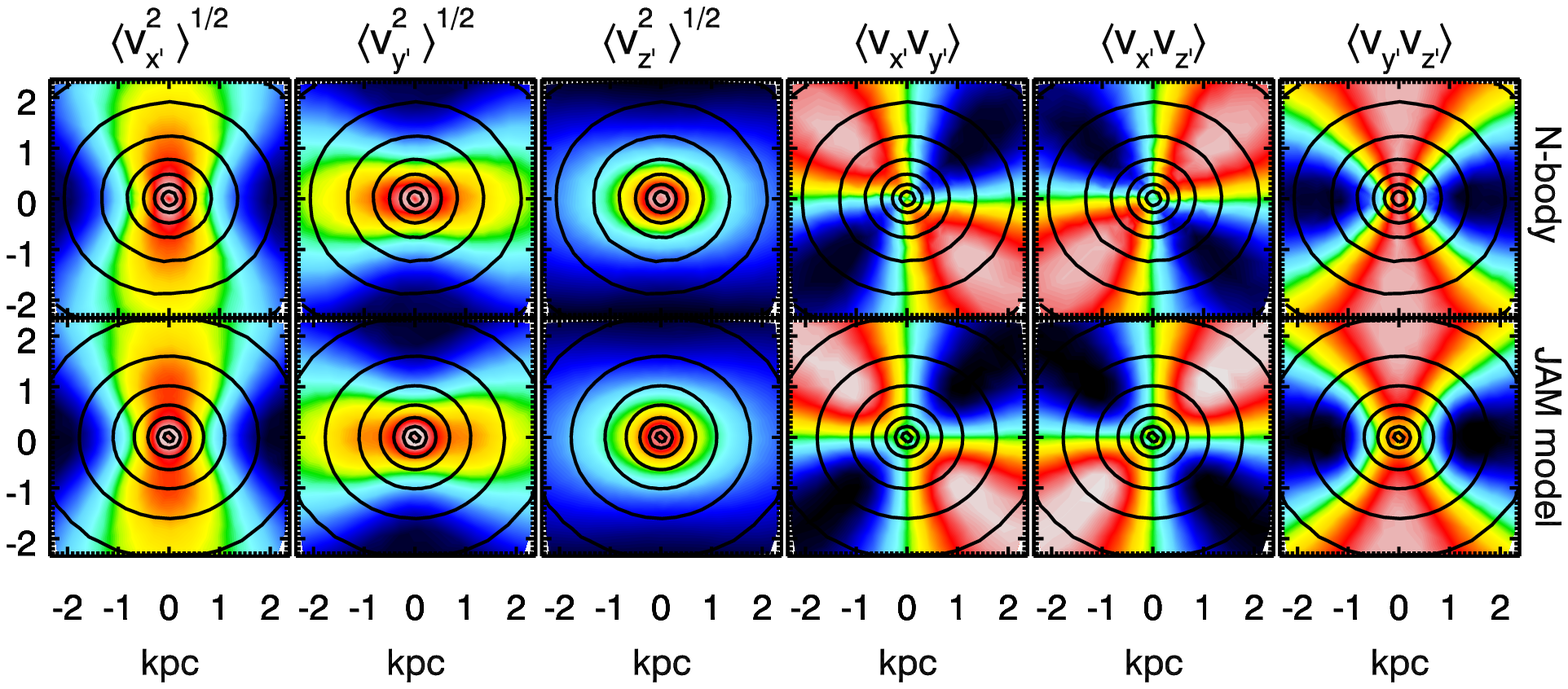}
  \caption{{\bf Projected second velocity moments of simulations versus JAM models.} {\em Top Row:} Voronoi-binned and linearly interpolated second moment calculated by direct summation (\refeq{eq:summation}) of the N-body particles of the evolved simulations N4179axi of \citet{Lablanche2012}. The simulation is projected at an inclination $i=60^\circ$. The corresponding velocity moment is written at the top of each image. {\em Second Row:} Same as in the top row for the second moments predicted by the JAM model (\refeq{eq:second_moment}) at the known inclination $i=60^\circ$. The MGE fit to the surface brightness of the above simulation was used as input to the JAM procedure. The anisotropy $\beta_z$ was fixed to the known average value $\beta_z^{\rm SIM}$ from the simulations.  The model is a prediction and not a fit to the data: it has no free parameters! {\em Third Row:} Same as in the top row, for the simulation N4570axi, projected at $i=30^\circ$. {\em Bottom Row:} Same as in the second row for the JAM model based on the MGE fitted to the simulation of N4570axi, projected at $i=30^\circ$. The simulations do not precisely satisfy the JAM assumptions, but the simple JAM models, with no free parameters, are able to provide a {\em striking} prediction for all six second moments, for both realistic N-body simulations.}
\label{fig:second_moments}
\end{figure*}

\section{Projected second moments}

In \citet{Cappellari2008} we used the \citet{Jeans1922} equations to derive the projected second velocity moments for an anisotropic (three-integral) axisymmetric stellar system with the density described via the Multi-Gaussian Expansion (MGE,  \citealt{Emsellem1994}). We stated in note 5 that the components of the proper motion dispersion tensor can be can be written via single quadratures without the need for special functions, and provided a reference software implementation, called the Jeans Anisotropic MGE (\textsc{JAM}) method\footnote{Available from http://purl.org/cappellari/idl}. However only the line-of-sight component was given in the paper. For completeness we provide all expressions in this addendum.

We adopt identical notation and coordinates system as in \citet{Cappellari2008}, and we refer the reader to that paper for details and definitions. Following the approach outlined in note 5 of the paper, we use the general formulas\footnote{With the substitution $x,y\rightarrow y,x$} (A5) of \citet{Evans1994} to write any of the six projected second moments as
\begin{eqnarray}\label{eq:second_moment}
    \lefteqn{\Sigma\,\overline{v_\alpha v_\beta}(x',y') = 4\pi^{3/2} G \int_0^1 \sum_{k=1}^N \sum_{j=1}^M\, \nu_{0k}\, q_j\, \rho_{0j}\, u^2\, \mathcal{F}_{\alpha\beta}} \nonumber \\
  &  & \times\frac{
\exp\left\{-\mathcal{A}\left[x'^2+y'^2 (\mathcal{A}+\mathcal{B})/\mathcal{E}\right]\right\}
  }{
  {\left(1-\mathcal{C} u^2\right) \sqrt{\mathcal{E}
       \left[1-(1-q_j^2)u^2\right]}}
  }\dd u, 
\end{eqnarray}
where $\alpha$ and $\beta$ stand for any of the three projected coordinates $x'$, $y'$ and $z'$, and we defined
\begin{equation}
    \mathcal{E}=\mathcal{A}+\mathcal{B}\cos^2 i.
\end{equation}

The line-of-sight velocity second moment $\overline{v_{\rm los}^2}\equiv\overline{v_{z'}v_{z'}}$ given in eq.~(28) of \citet{Cappellari2008} is obtained from \refeq{eq:second_moment} with
\begin{equation}
    \mathcal{F}_{z'z'}=\sigma_k^2 q_k^2 \left(\cos^2 i + b_k \sin^2 i\right) + \mathcal{D}\, x'^2\sin^2 i.
\end{equation}
The second moment $\overline{v_{x'}v_{x'}}$  of the proper motion in a direction parallel to the projected galaxy major axis is obtained with
\begin{equation}
    \mathcal{F}_{x'x'}=b_k \sigma_k^2 q_k^2 + \mathcal{D}\, \{[y'\cos i\, (\mathcal{A}+\mathcal{B})/\mathcal{E}]^2+\sin^2 i/(2\mathcal{E})\}.
\end{equation}
The second moment $\overline{v_{y'}v_{y'}}$ of the proper motion in a direction parallel to the the projected symmetry axis is obtained from $\mathcal{F}_{y'y'}=\mathcal{F}_{z'z'}(\pi/2-i)$, which implies
\begin{equation}
    \mathcal{F}_{y'y'}=\sigma_k^2 q_k^2 \left(\sin^2 i + b_k \cos^2 i\right) + \mathcal{D}\, x'^2\cos^2 i.
\end{equation}
The expressions for the cross terms are
\begin{equation}
    \mathcal{F}_{x'y'}=-\mathcal{D}\, x'y'\cos^2 i\, (\mathcal{A}+\mathcal{B})/\mathcal{E},
\end{equation}
\begin{equation}
    \mathcal{F}_{x'z'}=-\mathcal{F}_{x'y'}\tan i=\mathcal{D}\, x'y'\sin i\,\cos i\, (\mathcal{A}+\mathcal{B})/\mathcal{E},
\end{equation}
\begin{equation}
    \mathcal{F}_{y'z'}= \sin i\, \cos i\, [\sigma_k^2 q_k^2\, (1-b_k) - \mathcal{D}\, x'^2].
\end{equation}
The expressions for $\overline{v_{x'}v_{x'}}$ and $\overline{v_{y'}v_{y'}}$ where also recently given in \citet{DSouza2012}, who presented an application of the JAM method to proper motions measurements.

\section{Testing JAM against N-body simulations}

To test the formalism of the previous section we compared the six second moments predicted by the JAM method against the ones measured by direct summation from the N-body particles of two realistic galaxy simulations from \citet{Lablanche2012}, to which we refer the reader for details. Inside each projected pixels, with coordinates $(x'_l,y'_p)$ and $N$ particles, the second moments are
\begin{equation}\label{eq:summation}
    \overline{v_\alpha v_\beta}(x'_l,y'_p)=\frac{1}{N}\sum_{j=1}^N (v_{\alpha j}\, v_{\beta j}).
\end{equation}

To reduce the noise on the moments maps, we used the Voronoi binning method \citep{Cappellari2003} to group pixels in such a way that every bin contains about 5000 N-body particles. We then used the software of \citet{Cappellari2002mge} to fit an MGE model to the surface brightness of the simulation. The MGE parameters were used as input for the JAM model (\refeq{eq:second_moment}) to predict all six projected second moments. The model is not fitted to the data and it has no free parameters! We fixed the inclination $i$ to the known value and the anisotropy $\beta_z$ to the average values from the simulation particles (Table C1 of \citealt{Lablanche2012}). 

Considering the simplicity of the model assumptions and the lack of free parameters in our test, the agreement between the realistic simulations and the JAM model is {\em striking}, confirming the usefulness of the approach. This suggest that the method may be applicable to the interpretation of the observations from future generations of proper motions surveys  (generally using a maximum likelihood approach for discrete data). It can be especially useful to break the degeneracies that affect some of the quantities obtained from projected kinematics alone (e.g.\ the dark matter profiles).

\label{lastpage}

\end{document}